\documentclass[conference]{IEEEtran}
\IEEEoverridecommandlockouts
\usepackage{cite}
\usepackage{amsmath,amssymb,amsfonts}
\usepackage{algorithmic}
\usepackage{graphicx}
\usepackage{textcomp}
\usepackage{xcolor}
\usepackage[hyphens]{url}
\usepackage{booktabs}
\usepackage{multirow}
\usepackage{tabularx}
\usepackage{subcaption} 
\usepackage{comment}
\usepackage[english]{babel}  
\addto\extrasenglish{  
    
}
\usepackage{hyperref}  
\includecomment{SHOWN} 
\excludecomment{HIDDEN} 

\newcommand{\tv}[1]{\footnote{{\bf Thiemo: #1}}}

\newcommand{\Line}{\textsf{OBM}}
\newcommand{\Ring}{\textsf{Ring}}

\usepackage[table]{xcolor}   

\definecolor{cellgood}{RGB}{198,239,206}   
\definecolor{cellfair}{RGB}{255,222,170}   
\definecolor{cellbad}{RGB}{255,199,206}    
\newcommand{\good}{\cellcolor{cellgood}}
\newcommand{\fair}{\cellcolor{cellfair}}
\newcommand{\bad}{\cellcolor{cellbad}}

\usepackage[english]{babel}  
\addto\extrasenglish{  
    
}
\usepackage{tikz}
\usepackage{hyperref}  
\usetikzlibrary{arrows.meta, positioning, shapes, shapes.geometric, fit, backgrounds, calc, decorations.pathreplacing}

\def\BibTeX{{\rm B\kern-.05em{\sc i\kern-.025em b}\kern-.08em
    T\kern-.1667em\lower.7ex\hbox{E}\kern-.125emX}}
\begin{document}

\pdfpagewidth=8.5in
\pdfpageheight=11in

\newcommand{\iscasubmissionnumber}{971}

\pagenumbering{arabic}

\title{Broadcast, Don't Replicate: Eliminating Weight HBM in Datacenter LLM Serving}

\author{
    \IEEEauthorblockN{Hannah Atmer}
    \IEEEauthorblockA{
        Uppsala University \\
        Uppsala, Sweden \\
        hannah.atmer@it.uu.se
    }
\and
    \IEEEauthorblockN{Yuan Yao}
    \IEEEauthorblockA{
        Uppsala University \\
        Uppsala, Sweden \\
        yuan.yao@it.uu.se
    }
\and
    \IEEEauthorblockN{Thiemo Voigt}
    \IEEEauthorblockA{
        Uppsala University \\
        Uppsala, Sweden \\
        thiemo.voigt@angstrom.uu.se
    }
    \and
    \IEEEauthorblockN{Stefanos Kaxiras}
    \IEEEauthorblockA{
        Uppsala University \\
        Uppsala, Sweden \\
        stefanos.kaxiras@it.uu.se
    }
}

\maketitle
\thispagestyle{plain}
\pagestyle{plain}

\sloppy

\begin{abstract}
%

Large language model (LLM) inference at datacenter scale is fundamentally bottlenecked by memory capacity and bandwidth limits. Because model weights are read-only and identical across instances, conventional architectures that replicate these parameters across the local DRAM of tens of thousands of accelerators incur massive capital expenditures, significant static power dissipation, and redundant memory-fetch energy. A recent proposal to use optical fiber loops as delay-line memory alleviates these issues, but its recirculating loop must hold the whole model: loop length grows with model size, slack is fixed when the ring is built, and the layouts that serve modern Mixture of Experts (MoE) models efficiently do not fit in it.

We propose Optical Broadcast Memory (OBM), an open-ended hierarchical Tap-and-Amplify fiber architecture that generalizes and replaces the optical ring paradigm. A centralized memory server continuously broadcasts weights to distributed inference nodes that passively tap the data stream using receive-only co-packaged optics, delivering parameters directly to compute units without intermediate buffering. To overcome the limitations of continuous weight streaming, we use dynamically-sized slack to accommodate large and variable context lengths. We show how removing the restrictions of the optical ring allows OBM to efficiently serve MoE models. Modeled and simulated at a datacenter scale of 80,000 accelerators, OBM eliminates petabytes of HBM 
while being competitive or better in throughput and energy per token than an HBM-based system.

\end{abstract}

\section{Introduction}

Datacenter-scale LLM inference requires substantial memory for model parameters and KV caches. Unlike request-specific KV caches, model weights are read-only and identical across instances serving the same model. Nevertheless, conventional architectures replicate these weights across thousands of accelerators, multiplying storage costs, static power, and the dynamic energy of repeated memory reads to exactly the same data. This paradigm has driven a massive surge in DRAM/HBM demand, resulting in supply constraints, high costs, and thermal and power limits within hyperscale data centers~\cite{forbes}. Alongside power~\cite{IEA2026AI}, DRAM/HBM \emph{scarcity} is becoming one of the leading limiting factors in the deployment of LLMs.
We propose \emph{using the optical fiber network as shared weight memory, avoiding redundant storage and fetching}. Our approach rests on three insights:

\textbf{Insight 1: Weight replication wastes memory and energy.} Thousands of inference nodes store identical parameters and independently fetch them from local DRAM/HBM. Sharing a weight stream eliminates both replicated storage (cost) and redundant local reads (power).

\textbf{Insight 2: Optical broadcasting enables direct weight delivery.} A centralized transmitter broadcasts weights over a shared fiber network, which accelerators passively tap. Whereas GPUDirect RDMA~\cite{nvidia} delivers data into GPU memory and FPGA SmartNICs use intermediate buffers~\cite{fpgasurvey}, our approach streams weights directly to compute units, avoiding intermediate weight storage.

\textbf{Insight 3: Receive-only co-packaged optics simplifies integration.} Co-packaged Optics (CPO) places optical engines alongside compute silicon, shortening electrical paths and avoiding power-hungry transceivers and digital signal processing~\cite{fotouhi2019suma,HIRch9,HIRch8}. Silicon photonics (SiP) has gained significant traction for scaling ML clusters. Systems like SiP-ML~\cite{sip-ml} utilize CPO to replace electrical scale-up networks.
Our work re-imagines the asynchronous delay-line paradigm, replacing the fiber ring with an open-ended line topology. Our \emph{receive-only} approach also eliminates local lasers, reducing thermal and integration challenges compared with architectures such as Photonic Network-on-Wafer~\cite{photonic_chip}.

The concept of using an optical fiber ring as memory was recently proposed as a high-level vision to radically decrease the amount of DRAM needed for LLM inference~\cite{fiber}. In this bold proposal, akin to a delay-line memory~\cite{wiki:delay_line_memory}, immutable weights are injected into a continuous fiber ring; compute nodes passively tap into this loop, extracting the weights stream on the fly to achieve \emph{zero-copy} inference. However, this recirculating ring design suffers from fundamental physical limitations: the loop must hold the whole model, so its length (thousands of kilometers) grows with model size; its slack is built into the loop and cannot adapt to the load; and the layouts that serve Mixture of Experts (MoE) models efficiently, which repeat the weights every token uses on every cable, would make the loop longer still.

In this work, we generalize the Fiber Memory Ring approach and present \textbf{Optical Broadcast Memory (\Line{})}, the first optical network architecture designed to practically realize the optical shared memory paradigm. Instead of a recirculating ring (\Ring{}), a single centralized memory server continuously broadcasts the model weights on a \emph{hierarchical Tap-and-Amplify} topology. 

The architecture draws inspiration from early broadcast-driven architectures such as the Massive Memory Machine (MMM)~\cite{MMM} and DataScalar~\cite{datascalar}, which broadcast shared program data to distributed processors executing redundantly on local state. We extend these concepts to petabyte-scale LLM weight streaming.



In summary, in this paper we make the following contributions:
\begin{itemize}
    \item We propose Optical Broadcast Memory, an open-ended, hierarchical
    Tap-and-Amplify fiber line that streams weights through receive-only
    co-packaged optics directly into activation-stationary accelerators, without
    buffering (\autoref{sec:OBM}, \autoref{sec:hierarchy}).
    \item We formulate a roofline model for unbuffered weight streaming and show
    that in push-based architectures (\Line{}, \Ring{}) the compute knee is a
    hard limit (\autoref{sec:bandwidth}).
    \item We show that \Line{} breaks the \Ring{}'s fundamental limitation: a
    length that grows with the model it holds, which rules out replicating weights
    across cables and restricts the type of model parallelism that can be used. \Line{}'s
    length instead depends only on the accelerators it reaches, which frees them to
    use any execution model (\autoref{sec:execmodels}).
    \item We size attention slack per broadcast for the most loaded accelerator,
    with the amount of slack kept close to the mean by KV-aware request routing, and serve MoE models with a
    split-port data-layout 
    that keeps shared
    weights arriving at a rate below the knee of the roofline (\autoref{sec:slack}, \autoref{sec:evaluation}).
    \item We model \Line{} for an $80{,}000$-accelerator cluster serving
    Llama-3.1-405B, Llama-4-Maverick and DeepSeek-V4-Pro, and show that it removes
    $4.00$--$8.44$\,\textbf{petabytes} ($39$--$82\%$) of HBM while delivering
    $0.96$--$1.57\times$ the decode throughput of an HBM3e baseline at
    $10$--$42\%$ lower energy per token (\autoref{sec:evaluation}).
\end{itemize}

\begin{figure}
    \centering
    \includegraphics[width=1\linewidth]{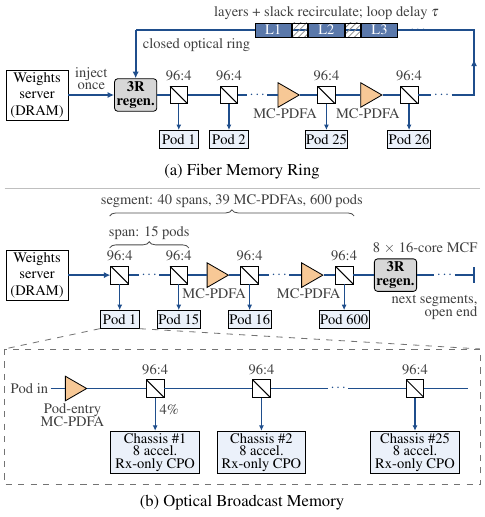}
    \caption{\Ring{} and \Line{} architectures}
    \label{fig:ring_obm}
\end{figure}

\section{Background: Fiber Memory Ring}




To eliminate redundant weight storage and local memory fetch energy, Atmer et al. recently proposed Fiber Memory~\cite{fiber}, an architecture that re-purposes optical fiber as an active, high-speed \emph{delay-line memory} for immutable LLM weights. Rather than storing weights at rest in local DRAM, a centralized weights server injects the model parameters into a shared optical fiber network, and inference accelerators passively tap the circulating light stream on the fly. The architecture eliminates massive amounts of DRAM from a datacenter and at the same time lowers the power needed to deliver weights to the accelerators.

To distribute the in-flight weight stream to tens of thousands of accelerators without incurring the latency and power overhead of traditional Optical-Electrical-Optical (O-E-O) packet switching~\cite{xue}, the baseline Fiber Memory architecture employs a hierarchical Ring-and-Pod topology (FM~\Ring{}) built around passive, asymmetric Tap-and-Amplify interfaces, as shown in~\autoref{fig:ring_obm}.a. At startup, a central weights server reads model parameters from HBM stacks and injects them into an optical ring, and in the original design an O-E-O 3R regenerator (Re-amplification, Re-shaping, Re-timing) renews the signal once per circulation~\cite{fiber}.

The compute node is an 8-accelerator baseboard chassis that ingests a 8-cable, 16-core MCF bundle. To avoid the manufacturing and thermal skew challenges of terminating 128 cores on a single die, the chassis routes exactly one MCF cable (providing data delivery bandwidth $3.2\text{ TB/s}$, comparable to the $3.35\text{ TB/s}$ HBM3 of an NVIDIA H100 SXM) to each of the 8 accelerators. Within each 8-accelerator chassis, the LLM is partitioned across an 8-stage pipeline synchronized with 8 time-shifted MCF core groups.
Accelerators follow an activation-stationary systolic execution model: dynamic activations and the KV cache reside in local HBM3e or on-chip SRAM, while the weights stream from the fiber, similar to the Cerebras Architecture~\cite{lie2023cerebras,lie2021multiwafer}.

Because the optical stream propagates continuously at $2/3 \text{ c}$ ($200,000\text{ km/s}$), an accelerator that stalls or experiences a prolonged memory lookup risks missing the start of the next layer, forcing it to wait an entire loop circulation ($\tau$). To prevent missed deadlines, the baseline architecture inserts optical padding (slack).
Unused slack causes accelerators to idle while insufficient slack causes missed deadlines. Either case translates into reduced throughput. Because the slack is part of the loop, a ring cannot adapt it to the load from one pass to the next; it must be sized, when the ring is built, for the worst load it will see. Maintaining phase alignment across 8 parallel MCF cables over a closed loop thousands of kilometers long exposes the system to significant thermal and mechanical length skew, requiring larger inter-layer slack margins.

\section{Optical Broadcast Memory}
\label{sec:OBM}

Our work, Optical Broadcast Memory (\Line{}), proposes an open-ended line, instead of a ring topology, as shown in~\autoref{fig:ring_obm}.b and detailed in~\autoref{sec:hierarchy}. 
Both topologies share the technical stack: 
weight broadcast, hierarchical optical tapping and amplification, receive-only co-packaged optics delivering weights directly to accelerator registers without intermediate buffering, 
but there are also significant differences.

In a ring topology the full model is read once in the weights server
and then re-circulated multiple times without active use of the memory in the weights server. Depending on the relative cost of reading the model out of the weights server and broadcasting it versus the cost of keeping the signal coherent in the ring via its regenerators, this could be an advantage of ring topologies that is forfeited when we ``sever the ring.'' In contrast, in an open-ended \Line{} architecture the full model is read from the weights server and broadcasted in consecutive passes.

On the other hand, the ring length is constrained by the need to fit the whole model plus allow sufficient space for slack to accommodate the increasing latency of ever expanding contexts and KV caches~\cite{fiber}. For large models, the ring length can become impractically large (depending of course on how much bandwidth is provided by the backbone ring). There is no such restriction in an open-ended broadcast \Line{}: It can be of any length, dictated solely by the need to accommodate the number of accelerators that feed off it. 
Lifting the ring length restrictions fundamentally changes the choices for the model parallelism that we can use (see~\autoref{sec:execmodels}) but also helps better manage thermal and length skew among the fiber cables (as they can be much shorter than the \Ring{}).



\section{Hierarchical Tap-and-Amplify}
\label{sec:hierarchy}

\begin{figure}[t]
    \centering
    \includegraphics[width=0.9\linewidth]{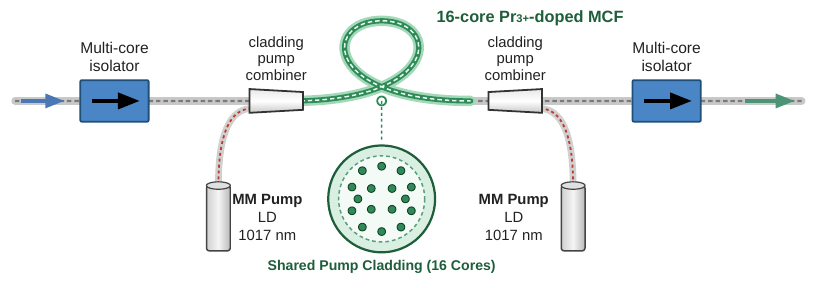}
    \caption{All-optical PDFA amplifier}
    \label{fig:pdfa}
\end{figure}

In an \Line{}, a weights server broadcasts the model weights on a fiber-optic backbone line (\autoref{fig:ring_obm}.b). The line consists of one or more
\emph{segments}, and each segment consists of several \emph{spans} connected by in-line all-optical PDFA amplifiers (\autoref{fig:pdfa}). 
Consecutive segments are joined by an O-E-O 3R regenerator, which is power-hungry because it converts the signal
to the electrical domain, processes it with DSP, and re-transmits it with
lasers~\cite{app10175912}. Adding segments extends OBM's accelerator capacity, but since a single
segment supports up to 120,000 accelerators (more than the 80,000 in our
evaluation) this work considers only single-segment configurations.
Within one segment, spans are connected with power-efficient \emph{all-optical} PDFA amplifiers, but there is a limit on how many such amplifications we can do before we need the expensive 3R regeneration.

\begin{figure}[h]
    \centering
\includegraphics[width=0.8\columnwidth]{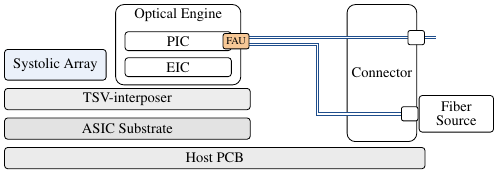}
    \caption{\emph{Receive-Only (no lasers)} Co-Packaged optics / accelerator integration. The fiber cable is coupled to the PIC through a fiber array unit (FAU). The Photonic and Electrical Integrated Circuits (PIC \& EIC) are positioned on the same interposer as the compute silicon 
    and feed raw weights directly into the systolic array via the interposer interface.}
    \label{fig:cpo}
\end{figure}



%
%
%
%
%
%


To maximize the number of accelerators we can reach within one segment, the key concept is a \emph{hierarchical, all-optical Tap-and-Amplify} approach. 
The optical signal is only converted to electrical at the accelerator's CPO receiver interface, mere millimeters away from compute where it is completely consumed (\autoref{fig:cpo}). 
Thus, the design goal is
to tap the signal as much as possible before amplifying it, and to keep the
loss along the path to every receiver small.
A passive tap divides signal and noise in the same ratio (96:4 in \autoref{fig:ring_obm}) and leaves the optical signal-to-noise ratio (OSNR) unchanged.
However, amplifiers that restore the signal strength as it degrades from tapping, add ASE (amplified spontaneous emission) noise in linear proportion to their gain~\cite{Andrekson}, which limits the number of spans per segment.
%
%

Consider a linear bus, where the signal reaching the last of $N$ nodes has crossed every
upstream tap. Each tap removes its tapped fraction and adds loss.
However, the number of taps that an amplifier can serve is limited and determined solely by the amplifier's output power and the sensitivity of the end-receivers, independently of the total $N$. 
A linear bus of $N$ nodes therefore
needs amplifiers at regular intervals, and the signal reaching its last node
passes through all of them. The number of amplifiers in that cascade, and the
noise they add, grows in proportion to $N$. 
Scaling a linear bus to more nodes thus degrades the OSNR of its farthest receivers in proportion to the number of nodes it serves.


A ``bus of buses'' breaks this dependence. In our design (\autoref{fig:ring_obm}), a backbone taps the
signal into $P$ pods, and within each pod a bus taps it into $n$ chassis, so
that $N = P n$. A signal now crosses at maximum $P$ backbone taps and $n$ pod taps ($P+n$), rather
than all $N$.
The
number of amplifiers in the backbone cascade, and the noise they add,
grows in proportion to the loss from the $P$ taps. A single amplifier restores the branch of the pod, setting a limit to the number of taps per pod.
While we only use two levels in this work (backbone and
pod buses), the same construction extends to a tree of arbitrary depth.
The total number of amplifiers remains proportional to $N$ in every topology.
The hierarchy changes only how many amplifiers each signal traverses. \autoref{fig:ring_obm} shows a full configuration of one segment with 40 spans (39 PDFA amplifiers), 15 pod taps per span (600 pods in total), and 25 chassis taps per pod. With 8 accelerators per chassis, this gives 120,000 accelerators. For the evaluation (\autoref{sec:evaluation}) we conservatively use 400 pods for 80,000 accelerators in total.



\section{Keeping up with streaming weights}
\label{sec:bandwidth}


The key enabler to stream weight parameters directly from the optical distribution network into the chip without intermediate weight buffering is co-packaged optics, shown in~\autoref{fig:cpo}. 
Because we stream LLM weights, which are a major memory bandwidth bottleneck during decode if read from memory~\cite{Pope2023, Aminabadi2022}, the accelerator utilizes an \emph{activation-stationary} systolic array~\cite{Sze2017}. 
Token activations are staged inside the processing elements beforehand, allowing the continuous broadcast weights to flow directly through the compute fabric without requiring edge weight buffers~\cite{lie2023cerebras,lie2021multiwafer}.
Only the weights---immutable, read-only data---arrive from fiber. Everything written during inference (token activations, partial sums and the
KV cache) stays in the accelerator's own memory: on-chip SRAM, its attached HBM stacks, or 
the HBM of another accelerator in the same chassis. 

\begin{figure}[t]
\centering
\includegraphics[width=0.8\columnwidth]{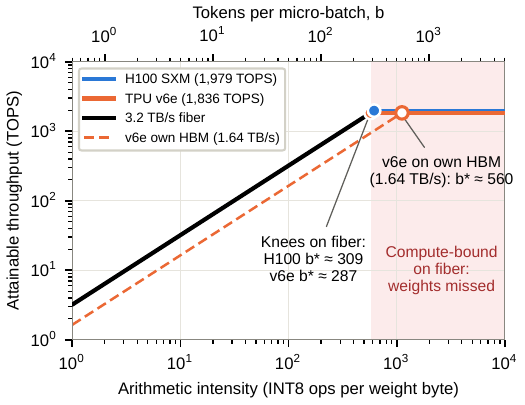}
\caption{Roofline of an NVIDIA H100 SXM and a Google TPU v6e, each fed by a
3.2~TB/s fiber stream, using published dense INT8/FP8 peak compute. Each weight
performs one multiply-accumulate (two operations) per token, so arithmetic
intensity is $2b$ operations per byte when each weight is applied to $b$
tokens (top axis). On fiber, both accelerators reach their compute roof near
$b^* \approx 300$; the TPU v6e on its own 1.64~TB/s HBM (dashed) needs
$b^* \approx 560$. Because the stream cannot be paused, operation to the
right of the knee is unreachable: compute falls behind and weights are
missed.}
\label{fig:roofline}
\end{figure}

The key insight for how \Line{} (and \Ring
) architectures work is that \emph{the weights that each accelerator receives must be consumed without the accelerator becoming compute-bound}.
For this, we turn to the roofline model~\cite{williams2009roofline}, which bounds an accelerator's
attainable throughput by the lesser of its peak compute $P$ and the product of
the bandwidth $B_a$ that feeds it and the arithmetic intensity $I$ of the
workload:
\begin{equation}
\Pi = \min\left(P,\; B_a \cdot I\right).
\label{eq:roofline}
\end{equation}
The two bounds meet at the knee, $I^* = P/B_a$. To the left of the knee the
accelerator is bandwidth-bound and part of its compute sits idle; to the right
it is compute-bound and additional bandwidth buys nothing.
\autoref{fig:roofline} shows the rooflines of a H100-class and a TPU v6e-class accelerator fed
at $B_a = 3.2$~TB/s, the bandwidth of one fiber-core group attached to the CPO of an accelerator, in our design.

\paragraph{Arithmetic intensity is set by the tokens per weight pass}
In \Line{} the accelerator is activation-stationary: it holds the
activation vectors of the tokens it is processing while the weights stream
past. We denote by $b$ the number of tokens per accelerator per weight pass,
that is, the number of tokens to which an accelerator applies each weight
during a single pass of the weights. Each weight streaming past is applied to
all $b$ activation vectors before it is gone, performing one
multiply-accumulate, or two operations, per token. With weights of $s_w$
bytes each, the arithmetic intensity is therefore $I = 2b/s_w$ operations per
byte, and the knee corresponds to
\begin{equation}
b^* = \frac{P \, s_w}{2\,B_a}
\label{eq:bstar}
\end{equation}
tokens per weight pass. Throughout this paper, $b$ is counted strictly per
accelerator and per weight pass. Choosing the fiber bandwidth per accelerator
is thus equivalent to choosing the largest $b$ the accelerator can serve. \autoref{tab:bstar_scaling}
summarizes how $b^*$ responds to each parameter. It grows with compute and
falls as bandwidth rises: doubling $B_a$ halves $b^*$, because weights arrive
twice as fast and each can be applied to only half as many tokens before the
next arrives. Weight precision matters less than it appears, because
accelerators typically trade halved weight size for doubled peak compute, and
the two effects cancel.

\begin{table}[t]
\centering
\caption{How the knee $b^* = P s_w / (2 B_a)$, in tokens per accelerator per weight pass, scales with peak
compute $P$, bandwidth per accelerator $B_a$ and bytes per weight $s_w$.}
\label{tab:bstar_scaling}
\begin{tabular}{lc}
\hline
\textbf{Change} & \textbf{Effect on $b^*$} \\
\hline
Peak compute $P \times k$                         & $b^* \times k$ \\
Bandwidth per accelerator $B_a \times k$          & $b^* \div k$ \\
8-bit $\rightarrow$ 16-bit weights, same $P$      & $b^* \times 2$ \\
8-bit $\rightarrow$ 16-bit weights, $P$ halved    & unchanged \\
\hline
\end{tabular}
\end{table}

\paragraph{Calibrating against current accelerators}
\autoref{tab:bstar_accel} applies \autoref{eq:bstar} to two current
accelerators, an NVIDIA H100 SXM and a Google TPU v6e, using their published
dense peak compute. NVIDIA specifies the H100's tensor throughput with 2:4
structured sparsity~\cite{nvidia_h100}, which doubles the dense rate per
streaming multiprocessor~\cite{choquette2023hopper}; we use the dense figures,
which are half the sparse ones. Microbenchmarks of the Hopper architecture
reach about 96\% of this dense peak in practice~\cite{luo2024hopper}. Google
lists the TPU v6e's peak compute without a sparsity mode~\cite{google_tpuv6e}.

\begin{table}[t]
\centering
\caption{Knee $b^*$, in tokens per accelerator per weight pass, for two accelerators, using published dense
peak compute, with their own HBM and with a 3.2~TB/s fiber stream.}
\label{tab:bstar_accel}
\setlength{\tabcolsep}{4pt}
\begin{tabular}{llrrrr}
\hline
 & & \textbf{Peak $P$} & \textbf{HBM} & \textbf{$b^*$} & \textbf{$b^*$} \\
\textbf{Accelerator} & \textbf{Prec.} & \textbf{(dense)} & \textbf{BW TB/s} & \textbf{(HBM)} & \textbf{(fiber)} \\
\hline
H100 SXM  & INT8/FP8 & 1,979 TOPS   & 3.35  & 295 & 309 \\
          & BF16     & 989 TFLOPS   & 3.35  & 295 & 309 \\
TPU v6e   & INT8     & 1,836 TOPS   & 1.64  & 560 & 287 \\
          & BF16     & 918 TFLOPS   & 1.64  & 560 & 287 \\
\hline
\end{tabular}
\end{table}

Two observations support our choice of 3.2~TB/s per accelerator. First, it
places the knee of both accelerators near 300 tokens per weight pass, which is
where the H100's own compute-to-HBM ratio places it. Second, it shows that
fiber bandwidth is not bound by the memory stacks on the package. With its own
HBM, the TPU v6e reaches its compute roof only at about 560 tokens per
weight pass, far beyond what decoding typically sustains. Fed at 3.2~TB/s from
fiber, it reaches the roof at about 290, half that size. Optical broadcast can
thus stream weights to each accelerator at the exact bandwidth its compute requires, without HBM bandwidth limiting throughput.

\paragraph{In a push system, the knee is a hard limit}
With HBM, a compute-bound accelerator simply runs slower: it fetches weights
when it is ready for them. Fiber pushes weights at a fixed rate and
nothing buffers them, so an accelerator whose compute falls behind the stream
\emph{misses weights} and must wait for the next broadcast pass 
to receive them.
The region to
the right of the knee, which a buffered memory could reach along the dashed
extension in \autoref{fig:roofline}, is therefore unreachable. The knee is
not merely the efficient operating point but the largest $b$ that
keeps pace with the stream. We therefore operate at $b < b^*$
that leaves a margin for compute jitter and for the gap between theoretical
and achievable peak; in \autoref{fig:roofline}, $b = 280$ leaves about
10\% and uses 91\% of peak compute. In the evaluation (\autoref{sec:evaluation}) we
derate peak compute to the measured GEMM efficiency $\eta = 0.85$, which lowers 
the hard limit to $b^* = 263$.

In this work, we focus on decode, which dominates serving time and is bound by weight
delivery. Prefill is compute-bound: a single prompt applies each weight to
hundreds or thousands of tokens, beyond the knee, so on \Line{} it must be
chunked to at most $b^*$ tokens per weight pass. At the knee, compute keeps
pace with the stream, so chunked prefill runs at about the same rate as
compute-bound prefill on HBM, but its tokens compete with decode tokens for
the same $b^*$ budget. We assume prefill is disaggregated~\cite{Zhong2024}, as
in production systems, and its KV transferred to the decode accelerators.
Integration of the disaggregated prefill in the \Line{} framework is left for future work.

\paragraph{Attention has its own roofline}
The roofline of \autoref{fig:roofline} covers the weight multiplications:
the attention projections, the output projection and the FFN. Attention over
the KV cache is different. It reads each request's own keys and values from
local DRAM, shares nothing across the $b$ tokens, since each token attends
to its own request's context, and has low arithmetic intensity whatever the
value of $b$; it is bounded by the DRAM bandwidth $B_\mathrm{DRAM}$, not by
the fiber. Its duration grows with $b$ and with the context length $C$. The variable duration of attention requires slack to be included in the optical broadcast. This is explained in \autoref{sec:slack}.

\paragraph{The tokens per weight pass are a design choice below $b^*$}
The knee bounds $b$ from above but, in practice, three constraints may hold it lower:
i) Timing jitter margin: compute must keep pace with a stream that cannot be
paused. ii) Attention latency: \autoref{eq:attnslack} ties $b$ to the
total context of the requests an accelerator holds, and through it to the
per-token latency (and, on a ring, to the loop capacity left for slack). iii) Supply: the
accelerator must have $b$ tokens ready at every weight pass. How much load the
system must carry to provide them depends on the execution model, which we
examine in \autoref{sec:execmodels}. When these constraints are met,
each accelerator applies every weight it receives to $b$ tokens, and the
weight-delivery energy per token falls in proportion to $1/b$, because the
same received bits serve $b$ tokens. 
\emph{Operating near the knee thus maximizes
both throughput and energy efficiency, and the fiber bandwidth per
accelerator is the parameter that places that knee where the accelerator's
compute can use it.}

\paragraph{Local memory and bandwidth}
Both \Ring{} and \Line{} relinquish massive bandwidth in the local accelerator memory (compared to a typical accelerator setup) by streaming immutable weights through fiber. Local memory, of course, handles all the read-write data: dynamic activations and the KV cache.  Reducing the bandwidth pressure to half on local memory, significantly reduces both the effective latency and the timing jitter for such accesses~\cite{esmailidokht2024mess}. 
Any fine-grain timing jitter that remains in these accesses can be hidden with prefetching and double-buffering inside the accelerator.

\begin{table*}[t]
\centering
\caption{Model parallelism on \Line{} and \Ring{}, for a chassis
of $N$ accelerators serving a model of $L$ layers and size $W$. The ring has
loop delay $\tau$, capacity $M$ and aggregate bandwidth $B$. All accelerators operate at a fixed input bandwidth $B_a$ and are not compute bound (see~\autoref{sec:bandwidth}). Each accelerator applies every weight it receives to $b$ tokens, and the chassis serves $R$ concurrent requests.
Shading: green is favorable, orange intermediate, red unfavorable relative to the other execution models.}
\label{tab:execmodels}
\begin{tabular}{lccc}
\hline
 & \textbf{Pipeline} & \textbf{Data-parallel} & \textbf{Tensor-parallel} \\
\hline
Weights per accelerator             & $L/N$ layers          & All $L$ layers         & $1/N$ of every layer \\
Distinct weight stream              & $B$ ($N$ groups) & $B/N$, split $N$ ways & $B$ ($N$ groups) \\
Chassis throughput                  & \good $b$ tokens per $\tau$ & \good $b$ tokens per $\tau$ & \good $\approx$ $b$ tokens per $\tau$ \\
Per-token latency                   & \bad $N\tau$          & \bad $N\tau$           & \good $\approx\tau$ + all-reduces \\
Requests per chassis $R$            & \fair $N b$            & \fair $N b$           & \good $b$ \\
Requests to reach the knee          & \fair $N b^*$          & \fair $N b^*$           & \good $b^*$ \\
KV cache per accelerator            & \bad $L/N$ layers $\times$ $N b$ req. & \bad $L$ layers $\times$ $b$ req. & \good $1/N$ KV heads, $L$ layers $\times$ $b$ req. \\
Inter-accelerator traffic per token & \fair $N-1$ vectors   & \good None             & \bad $4L(N-1)$ vectors \\
Cross-cable synchronization         & \fair Once per stage  & \good None             & \bad Twice per layer \\
Synchronization points per token    & \fair $N-1$ handovers & \good None             & \bad $2L$ all-reduces \\
\hline
\textbf{Fiber Memory---Ring}        & \good Feasible~\cite{fiber} & \bad Infeasible ($M/N < W$) & \fair Needs per-layer cross-cable deskew \\
\textbf{Optical Broadcast Memory---\Line{}} & \good Feasible & \good Feasible & \good Feasible \\
\hline
\end{tabular}
\end{table*}

\section{Model Parallelism for the \Ring{} and \Line{}}
\label{sec:execmodels}

\begin{figure}
    \centering
    \includegraphics[width=1\linewidth]{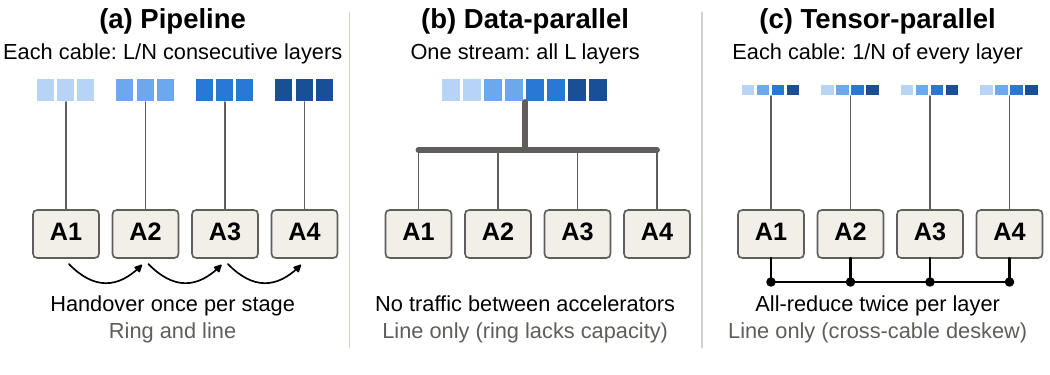}
    \caption{\Line{} and \Ring{} execution models.}
    \label{fig:exec_models}
\end{figure}

Section~\ref{sec:bandwidth} characterized each accelerator by $b$, the
number of tokens to which it applies each weight during one pass of the
weights. A chassis, however, is loaded by requests. Let $R$ be the number of
concurrent requests a chassis serves. In decode, a request contributes one
token per step and cannot start its next token before the current one has
passed through all $L$ layers. How $R$ maps onto $b$ therefore depends on how
the execution model divides the layers among the accelerators. This mapping,
together with how often the accelerators must synchronize, distinguishes the
three primary execution models: pipeline (PP), data-parallel (DP), and tensor-parallel (TP), shown in \autoref{fig:exec_models} and compared in \autoref{tab:execmodels}.
Expert parallelism (EP) is handled in the evaluation (\autoref{sec:evaluation}) where it is used in Mixture of Experts (MoE) models.

\subsection{\Ring{}: pipeline and tensor parallelism.}

Consider a chassis of $N$ accelerators serving a model of $L$ layers and size
$W$. Each accelerator taps one of $N$ fiber-core groups, each carried in its own MCF
cable. The ring has loop delay $\tau$, in-flight capacity $M$ and aggregate
bandwidth $B$, so each fiber-core group holds $M/N$ and delivers $B/N$ of data.

As shown in~\cite{fiber}, in the pipeline, each accelerator executes a contiguous block of $L/N$ layers,
and every weight it consumes arrives on the same cable. Each of the $N$ stages
works on a different group of $b$ requests, and a request's next token cannot
enter the first stage until its current token has left the last. Keeping
every stage busy therefore requires $R = N b$ concurrent requests per
chassis. The only skew it must
tolerate is between the cores of that cable, which share a cladding and a
thermal environment; this skew is very small ($<$1.5\,ps~\cite{Puttnam}).
Skew between cables matters only when a pipeline stage hands its
activations to the next, once per loop. Over long distances, temperature
changes make the cables drift relative to each other: fiber delay changes by
about 39~ps/(km$\cdot$K)~\cite{slavik2015thermal}, so the drift grows in
proportion to the ring's length---and the rings can be large (thousands of km) to provide sufficient capacity. 
Drift can be corrected only where the signal is retimed, at an O-E-O 3R
regenerator. The original design places a single 3R regenerator per
loop~\cite{fiber}; noise accumulation in fact requires one about every
450\,km (\autoref{sec:architecture}), so a ring of thousands of kilometers needs
about ten, each of which must also deskew the 8 cables.
With sufficient deskew buffering, such drift is manageable at one
synchronization point per loop and could be hidden by introducing inter-layer slack at stage boundaries~\cite{fiber}. Consecutive layers within a stage arrive on the same cable and need no slack for cable skew.

The ring rules out data parallelism, 
in which every
accelerator receives the full model, because it does not fit: a single fiber-core group holds
only $M/N$ of the weights, a fraction of the model and weight replication across fiber-core groups is constrained by the need to fit the whole model on the aggregate capacity of all fiber cores  (\autoref{tab:execmodels}). Tensor parallelism, however, fits, since each core
group could carry $1/N$ of every layer, but it requires all $N$ cables to be
aligned within every layer, with two all-reduce operations per layer. That means
cross-cable synchronization $2L$ times per token rather than once per stage,
on the longest cables in the system. Cross-cable timing drift (skew) may require deskew buffers on the ring. While in our evaluation we assume that skew is tracked and corrected at no cost, we also quantify throughput cost when it is not. However, we leave the hardware cost of the deskew to future work.

\subsection{\Line{}: the length constraint disappears; all models viable}
An \Line{} need only be long enough to reach its accelerators, which
is typically orders of magnitude shorter 
than a ring sized to hold the model.
Inter-cable skew shrinks by the same factor, and all three alternatives become
practical (\autoref{tab:execmodels}).

Data parallelism needs only one distinct weight stream. Since every
accelerator receives the full model, all $N$ core groups would carry identical
data.
Per-token latency is unchanged,
because a token must still see all $L$ layers stream past at $B/N$, and throughput matches the pipeline, but there is no handover between accelerators at all. Each accelerator serves its own $b$ requests from the first layer to the last, so the chassis again needs $R = N b$ concurrent requests. Each accelerator holds the KV cache for all $L$ layers, but only for its own $b$
requests, whereas a pipeline stage holds $L/N$ layers for all $N b$ requests in flight. At equal $R$, the KV cache per accelerator is therefore the same in both models.

Tensor parallelism maps naturally onto grouped-query attention models whose
number of KV-head groups is a multiple of $N$: each accelerator computes
attention for its share of the groups and $1/N$ of each FFN. All $N$
accelerators apply their slices of the weights to the same $b$ tokens, so
$R = b$: tensor parallelism gives each accelerator $b$ tokens per weight pass
with $N$ times fewer concurrent requests than the pipeline or data
parallelism. Both the weight
stream and the attention slack of each layer shrink by a factor of $N$,
reducing per-token latency by roughly $N$,
plus the cost of the all-reduces.
It is the only one of the three models that improves latency rather than
throughput, and it is practical only where cross-cable skew is small, as on a short \Line{}; on a ring it needs per-layer deskew (\autoref{tab:execmodels}).

The three models also differ in how much the accelerators of a chassis must
communicate. Let $d$ be the model's hidden dimension. In the pipeline, each of
the $N-1$ stage handovers transfers one $d$-element activation vector,
point-to-point, so a token moves $N-1$ vectors and synchronizes $N-1$ times.
In data parallelism, each accelerator executes all $L$ layers itself and
exchanges nothing per token. Tensor parallelism requires two all-reduce
operations per layer, after attention and after the FFN; with a ring
all-reduce, each moves $2(N-1)/N$ vectors per accelerator, so a token moves
$4L(N-1)$ vectors in total and synchronizes $2L$ times. That is $4L$ times the
pipeline's traffic, on the critical path of every layer. The KV cache adds no
traffic in any model, since it stays with the accelerator that computes the
corresponding attention. All volumes are per token; per weight pass, they
scale with $b$.


\section{Slack and KV Load Balancing}
\label{sec:slack}

\begin{figure}[t]
    \centering
    \includegraphics[width=0.8\linewidth]{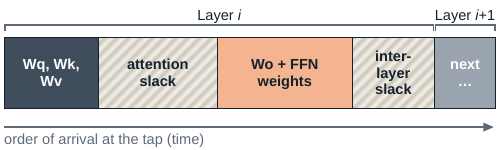}
    \caption{Attention and inter-layer slack}
   \label{fig:slack}
\end{figure}

Slack modulates weight delivery to allow accelerators to finalize their activation memory swaps before the next layer payload begins and, more importantly, to accommodate the increasing latency of attention with increasing context size (\autoref{fig:slack}). 
%
Attention grows with the sum of all context lengths $C_i, \quad i = 1, \dots, b$
for a batch size $b$. The KV cache access latency scales predictably with sequence length~\cite{lyceum_kv_cache_memory}, so the weights server 
can allocate an amount of space between attention and  FFN weights to accommodate a given context length.
If each token adds $s_{kv}$ bytes per layer to the KV cache held by the
accelerator, the attention slack of each layer
must satisfy
\begin{equation}
t_\mathrm{slack} \;\geq\; \frac{\, s_{kv} \sum_{i=1}^{b} C_i }{B_\mathrm{DRAM}} .
\label{eq:attnslack}
\end{equation}
A larger $b$ therefore makes the weight stream more efficient but
lengthens the attention slack, which 
increases token latency.


On a lockstep line or ring, the attention slack of a layer must cover the
attention work of \emph{every unit that shares the broadcast}. Decode attention is
bound by memory bandwidth, so that work depends on the total context a unit
holds, not on its longest request. For a unit $u$ holding the KV of requests
$\mathcal{R}_u$, \autoref{eq:attnslack} becomes
\begin{equation}
t_{\mathrm{slack}} \;\ge\; \max_{u}\; \frac{s_{kv}}{\theta\,B_{\mathrm{DRAM}}}
\sum_{i\in\mathcal{R}_u} C_i ,
\label{eq:slacksum}
\end{equation}
where the unit is an accelerator under DP and PP, and a chassis under TP, whose
accelerators each read $1/\theta = 1/N$ of every request's KV ($\theta = 1$ when
the model has fewer KV heads than $N$ and its KV is replicated).
A global scheduler keeps every accelerator’s total context under a fixed budget, and the server sizes the slack for that budget. 
The placement policy is already in research and production LLM schedulers~\cite{llumnix,aladdin,mooncake,dynamo_router}: each new request goes to the unit with the
least current KV load.
A request with a $32$k-token context, the longest we allow, is placed on
an accelerator with room for it; it takes about $32$k tokens of that
accelerator's budget, so the accelerator holds fewer other requests.
No accelerator ever exceeds the budget, so none misses the FFN weights, and the slack is set by the budget rather than by the longest request. 


\section{\Line{} Optical Network}
\label{sec:architecture}

\textbf{MCF and Amplifiers:}
The \Line{} is built with 8 MCF cables, each containing 16 cores with 16 wavelengths/core at 100 Gbps each for a total bandwidth of 25.6 TB/s across the 8 cables. The full network bandwidth enters the chassis via all 8 MCF cables, and each accelerator receives the fiber optic cores split out from one MCF cable. To amplify all cores in a MCF cable with a single device, we copy the conceptual design of a multi-core PDFA (MC-PDFA) described by Atmer et al.~\cite{fiber}. A standard O-band PDFA produces a signal gain of 25-30 dB~\cite{fiberlabs_pdfa}. The MC-PDFA is conservatively assumed to provide a $+20\text{ dB}$ gain based on a multi-core EDFA described by Zhang et al.~\cite{Zhang2024_MCFA}.

\textbf{Taps and Loss budgets:}
The pods and the chassis tap the main line and the pod bus respectively with a 96:4 passive splitter, where 4\% of the signal is extracted for a theoretical loss of $10\log_{10}(0.04)\approx 14$ dB loss on the tap line and a theoretical $10\log_{10}(0.96)\approx 0.2$ dB loss on the through line~\cite{fiber}. There is also additional excess loss where power is wasted as heat or scattered within the coupler. The weights server laser generates +4 dBm~\cite{laser_source}. The PAM4 receiver on the accelerator needs at least -14dBm to read the signal. Because every receiver sits behind a 4\% tap, a signal launched at $P_\mathrm{out}$ dBm can lose at most $P_\mathrm{out} - 14 + 14 = P_\mathrm{out}$ dB along a through line before the last receiver falls below sensitivity: only 4\,dB directly behind the +4\,dBm server laser, or about 20 taps at 0.2\,dB, and more behind a pod booster with a higher output power. The number of taps on the main line and in a pod is determined by the loss budget that we have from the weights server or after the signal is amplified or regenerated.

\textbf{Noise Budget and Regeneration:}
The MC-PDFAs add ASE noise to the signal during amplification. The signal must be regenerated by a 3R regenerator before it becomes too noisy. An MC-PDFA is assumed to add 6 dB of noise to the signal, since this is a typical amount for a regular PDFA~\cite{pdfa_noise}. An amplifier is placed at the pod entrance after the tap to increase the strength of the signal. Amplification creates noise, so keeping as many amplifiers off the main line as possible optimizes the cumulative noise on the main line.

PAM4 modulation requires a minimum OSNR of $\approx$ 29 dB to read the signal~\cite{pam}, which arrives at the accelerator only after a 4\% tap off the main line, amplification at the pod entry, and a final 4\% tap off the pod bus. Because the power drops after every tap, the final tap before a main-line MC-PDFA has the lowest signal power. When 4\% of the already-attenuated signal is diverted into the pod entry MC-PDFA, the input power has dropped by a massive 13.98 dB relative to the main line. To survive this, the pod entry MC-PDFA generates so much noise that it is mathematically equivalent to the noise from about 25 main-line MC-PDFAs. Therefore we choose network parameters that ensure that the worst-case OSNR stays above 29 dB for the final tap on the segment.

Finding the optimal number of taps and MC-PDFAs per segment requires balancing the linear insertion loss and logarithmic ASE noise accumulation. The number of MC-PDFAs, and the number of taps per span or pod bus that can be placed on the main bus or a pod bus, before needing to regenerate the signal with a 3R regenerator, is calculated for the case study by applying a Pareto filter and finding the "knee" of the Pareto curve showing the number of taps that can be placed per amplifier. 

The original \Ring{} design~\cite{fiber} placed an inline $+20\text{ dB}$ MC-PDFA after each $50\text{ km}$ spool to offset $16.0\text{ dB}$ of O-band fiber attenuation ($0.32\text{ dB/km}$~\cite{Rademacher2020}) plus $4.0\text{ dB}$ of passive tap/insertion loss, claiming a single central 3R regenerator suffices for the entire ring. However, that analysis balanced only optical power while overlooking ASE noise accumulation and chromatic dispersion. A pod tapped at the end of a spool, where the signal has already lost $20\text{ dB}$, drops by another $14\text{ dB}$ across the $4\%$ pod tap, so its booster sees about $-30\text{ dBm}$; with a $6\text{ dB}$ PDFA noise figure, the OSNR then falls below $24\text{ dB}$ after a single $50\text{ km}$ spool, violating the $29\text{ dB}$ minimum OSNR required for $100\text{ Gbps}$ PAM4 modulation. The \Ring{} must therefore tap its pods right after each inline amplifier. Even then, each spool's amplifier contributes an OSNR of about $39\text{ dB}$, and the noise of the cascade limits a ring to 9 spools ($450\text{ km}$) between O-E-O 3R regenerators (\autoref{sec:evaluation}), not one regenerator per loop.

\section{Evaluation}
\label{sec:evaluation}

We evaluate Optical Broadcast Memory (\Line{}) against a datacenter-scale HBM3e baseline and the recirculating Fiber Memory Ring (\Ring{})~\cite{fiber} on an $80{,}000$-accelerator cluster serving three models: a dense model (Llama-3.1-405B), a coarse-grained MoE (Llama-4-Maverick)~\cite{maverick}, and a fine-grained MoE with compressed sparse attention (DeepSeek-V4-Pro~\cite{deepseekv4}). We report the combined static and dynamic energy cost per token, cluster throughput, and per-token decode latency. 

\subsection{Methodology}

\paragraph{Cluster and accelerators}

We model a datacenter cluster of $80,000$ H100-compute-class accelerators organized into $10,000$ 8-accelerator HGX-style chassis ($N=8$, grouped into $400$ pods of $25$ chassis). Each accelerator features $P = 1,979\text{ TOPS}$ of dense FP8 tensor compute~\cite{nvidia_h100}, of which a fraction $\eta = 0.85$ is achieved on weight GEMMs in every system~\cite{luo2024hopper}, provisioned with $128\text{ GB}$ of local HBM3e operating at $B_{\mathrm{DRAM}} = 3.35\text{ TB/s}$. For \Ring{} and \Line{}, each chassis ingests an 8-cable, 16-core, 16-wavelength/core ($100\text{ Gbps}/\lambda$) MCF bundle ($25.6\text{ TB/s}$ aggregate trunk bandwidth per chassis), delivering an aggregate bandwidth of $3.2\text{ TB/s}$ to each accelerator via receive-only co-packaged optics. Accelerators running TP exchange activations over NVLink at an all-reduce bus bandwidth of $400$\,GB/s~\cite{nvidia2018nvswitch} with $8$\,$\mu$s per collective.

\paragraph{Model Specifications}
\autoref{tab:eval_models} describes the three evaluated models.

\begin{table}[th]
\centering
\setlength{\tabcolsep}{2pt}
\begin{tabular}{lrrr}
\toprule
 & \textbf{Llama-3.1} & \textbf{Llama-4} & \textbf{DeepSeek-} \\
 & \textbf{405B} & \textbf{Maverick} & \textbf{V4-Pro} \\
\midrule
Total / Active Param. & $405\text{B} \; / \; 405\text{B}$ & $400\text{B} \; / \; 17\text{B}$ & $1.6\text{T} \; / \; 49\text{B}$ \\
Weight Precision & FP8 & FP8 & FP8 \& MXFP4 \\
Weights Size (GB) & $405$ & $400$ & $845$ \\
Layers (Dense/MoE) & $126/0$ & $24/24$ & $0/61$ \\
Hidden Dim. & $16,384$ & $5,120$ & $7,168$ \\
Attention Type & GQA, 8 KV & GQA, 8 KV & MLA + CSA/HCA \\
Attention per Layer & $4,096\text{ B}$ & $4,096\text{ B}$ &$1,152\text{ B}$ \\
KV per Token & $516\text{ KB}$ & $197\text{ KB}$ & $6.0\text{ KB}$ \\
Routed experts / top-$k$ & -- & 128 / 1 & 384 / 6 \\
Shared / Experts (GB) & $405 \; / \; 0$ & $16 \; / \; 384$ & $26 \; / \; 819$ \\
Baseline weights/GPU & 50.6\,GB & 50.0\,GB & 105.5\,GB \\
\bottomrule
\end{tabular}
\caption{Evaluated models. KV is per token of context; DeepSeek-V4-Pro also holds a 128-entry sliding window and compressor state per request.}
\label{tab:eval_models}
\end{table}

\paragraph{Systems and execution models}
Each system is evaluated with the execution models of \autoref{tab:execmodels}:
    \textbf{\textsf{Baseline} (HBM3e).} Each GPU stores its weight shard. We
    use TP for attention and dense layers,
    with experts partitioned across GPUs (TP+EP). Because V4-Pro has a single KV
    head, this plan must replicate its KV on all 8 GPUs. A step is
    \begin{equation}
    T = \max\!\Big(\sum_{s} \max\big(T_{w,s},\, T_{\mathrm{mm},s}\big),\;
        T_{\mathrm{coll}}\Big) + \tfrac{1}{16}T_{\mathrm{coll}} + A ,
    \label{eq:base_step}
    \end{equation}
    where the sum runs over two weight segments, the weights every token uses
    (attention, dense and shared layers) and the routed experts. $T_{w,s}$ reads a
    segment's weights from HBM (for experts, only those its tokens touch),
    $T_{\mathrm{mm},s}$ is its GEMM time at $\eta P$, $T_{\mathrm{coll}}$ the
    all-reduces or all-to-alls, and $A$ the attention time for the KV the GPU
    actually holds. Micro-batching overlaps collectives with compute, and we leave
    $1/16$ of their time exposed.

    \textbf{\Ring{} (PP$=$8).} One ring per model, each stage streaming its
    $L/8$ layers through one port~\cite{fiber}, plus seven activation handovers per
    token. One loop is one pipeline stage.
    
    \textbf{\Ring{} (TP).} Each of the 8 cables carries an eighth of every
    layer, and for the MoE models an eighth of every layer's experts (TP+EP), so no
    weight is repeated across cables. One loop is one decode step. The 8 cables must
    agree at every all-reduce, $2L$ times per token; we assume that cross-cable
    skew is tracked and corrected, for example at the 3R stations, and give its cost
    when it is not in the sensitivity analysis. A ring cannot use DP-attention with
    EP, which repeats the weights every token uses on all 8 cables: the loop would have to hold those weights 8 times, which for Maverick lengthens it to $8{,}300$\,km, $2.2\times$ its PP ring. We do not evaluate this layout.

    Each ring is sized for its execution model and operating point: its loop holds one eighth of the weights plus the attention slack of one loop, sized for $1.035\times$ (PP) or $1.04\times$ (TP) the mean KV load. Under PP, Llama-3.1-405B needs $3{,}718\text{ km}$ ($\tau = 18.6\text{ ms}$), Llama-4-Maverick $3{,}676\text{ km}$ ($18.4\text{ ms}$) and DeepSeek-V4-Pro $6{,}640\text{ km}$ ($33.2\text{ ms}$); under TP, $4{,}613\text{ km}$ ($23.0\text{ ms}$), about $3{,}780\text{ km}$ ($18.9\text{ ms}$) and about $6{,}820\text{ km}$ ($34.1\text{ ms}$), in $50\text{ km}$ spools.

    Each spool ends in an inline MC-PDFA, and the ring taps at most 16 pods right after each amplifier, for 400 pods with one MC-PDFA each. Noise accumulation requires an O-E-O 3R regenerator every 9 spools ($450\text{ km}$; see Power and energy below), so the rings need 9 to 16 3R stations, which also retime and deskew the signal.
    Each station adds under $2\,\mu$s of latency~\cite{Yin2022FEC}, which is negligible next to $\tau$, even for a large number of revolutions. Although in \Ring{} the 3R latency accumulates with each revolution, it does not affect \Line{}.

    \textbf{\Line{} (DP).} One line per model; every accelerator ingests the
    full model. For the MoE models this includes every expert: its $b\approx 250$ tokens need almost every expert in every MoE layer.
    
    \textbf{\Line{} (TP) and \Line{} (DP-attn.\,+\,EP).}
    The 8 ports of a chassis carry different weights, in one of two layouts, both
    reported. Under TP, each port carries an eighth of every layer, and for the MoE
    models an eighth of every layer's experts (TP+EP), as on the TP \Ring{}. Under
    DP-attention with EP, which applies only to the MoE models, each port carries
    the weights every token uses, identical on all 8 cables, plus its eighth of the
    routed experts; each accelerator runs attention for its own requests and sends
    tokens to the accelerator holding their experts. All-reduces and all-to-alls overlap the
    stream when the weights server orders each row-parallel matrix by
    output-column blocks and the following matrix by input-row blocks: each
    block's collective then runs while the next block streams. We charge the same
    $1/16$ exposed time as the baseline.
    The \Line{} requires only one fiber optic cable segment to serve all 80,000 accelerators, with 400 pods of 25 chassis each. Each pod contains one MC-PDFA, and the main line needs 26 MC-PDFAs, for a total of 426 MC-PDFAs.
    
    Finally, we call a layout \textbf{split-port} if it assigns different slices of each layer to the 8 ports of a chassis (TP, DP-attn.\,+\,EP for MoE), so that a decode step waits for only $1/8$ of the model's stream, and a \textbf{whole-model stream} (\Ring{} PP, \Line{} DP) if each decode step waits for the whole model.
\begin{HIDDEN}
    \item \textbf{\Line{} (DP).} Every accelerator ingests the
    full model. 
    \item \textbf{\Line{}}
    For DP, every accelerator ingests the full model. 
    For TP and DP+EP, The 8 accelerators of a chassis carry different weights, in one of two layouts. Under TP, each port carries an eighth of every layer, and for the MoE
    models an eighth of every layer's experts (TP+EP), as on the TP \Ring{}. 
    For expert layers, the \Line{} uses expert parallelism, where each accelerator in the chassis receives $1/8\text{th}$ of the experts. For dense layers, each of the eight MCF cables carries the full model, while for expert layers the experts are split across the cables so that each accelerator receives its own distinct set of experts. Under DP-attention with EP, each accelerator runs attention for its own requests and sends
    tokens to the accelerator holding their required experts. All-reduces and all-to-alls overlap the
    weight stream when the weights server orders each row-parallel matrix by
    output-column blocks and the following matrix by input-row blocks: each
    block's collective then runs while the next block streams. We charge the same
    $1/16$ exposed time as the baseline. The \Line{} requires only one fiber optic cable segment to serve all 80,000 accelerators, with 400 pods of 25 chassis each. Each pod contains one MC-PDFA, and the main line needs 26 MC-PDFAs, for a total of 426 MC-PDFAs.
\end{HIDDEN}

\paragraph{Operating point}
Following \autoref{tab:execmodels}, a chassis serving $R$ requests applies each
weight to $b = R/N$ tokens per pass under DP and PP, and to $b = R$ under TP.
Rather than fixing $b$, we run every system at the knee of its own throughput curve. For each system, we find the highest throughput per accelerator it can reach with up to $2{,}048$ requests per chassis, and then pick the smallest load that comes within $3\%$ of that peak. Adding requests beyond this point barely raises throughput and only lengthens time between tokens (TBT).

On \Ring{} and \Line{} the knee of \autoref{sec:bandwidth} is also a hard limit,
and it applies to every weight byte as it arrives, not to the pass on average.
The weights every token uses are applied to all $b$ tokens of an accelerator, so
$b$ may not exceed $b^* = \eta P s_w/(2B_a) = 263$ tokens, however many experts
the model has. Routed experts are applied only to the $\approx b k/E$ tokens routed
to them, far below the knee, but the time they leave idle cannot be lent to the
dense weights, which have already passed. We therefore charge each weight segment
$\max(\text{stream}, \text{compute})$ separately and reject operating points
where any segment's compute outpaces its stream. The baseline fetches weights
when it is ready for them, so for it the knee only marks where it becomes
compute-bound. In this way DP and PP reach their knee near $R = N b^*$, and TP
near $R = b^*$, as \autoref{tab:execmodels} predicts, 
for MoE and dense models.

\paragraph{Load-based placement and per-broadcast slack}

All three systems use the same placement policy, as production LLM schedulers
do~\cite{llumnix,aladdin,mooncake}: each new request goes to the unit with the
least current KV load, and admission reserves KV for $1.05\times$ the mean load.
We simulated this policy over $2{,}000$ decode steps (two seeds), with every
context growing by one token per step and finished requests replaced by new ones.
On average, the most loaded of the $10{,}000$ chassis or $80{,}000$ accelerators
carries $1.010$--$1.015\times$ the mean load. In the worst step it carries
$1.017\times$ on the MoE models' accelerators, $1.029\times$ on 405B's accelerators
and $1.037\times$ on 405B's TP chassis. The excess stays within about one request:
the policy sends the largest new requests to the least loaded units, so the worst
case is a $32$k-token request, the clip of the distribution, landing on a unit just
below the mean. Such a request is $9\%$ of a TP chassis's load at $R = 248$.
Random placement would give $1.6$--$1.9\times$, and power-of-two
choices~\cite{mitzenmacher2001power} $1.13$--$1.26\times$.
\begin{itemize}
    \item \textbf{\Line{}:} the weights server sets each broadcast's slack for the
    most loaded unit, which we budget at $1.02\times$ the mean.
    \item \textbf{\Ring{}:} the slack is built into the ring's length, so it must
    cover the worst case: $1.035\times$ the mean on the PP ring's accelerators and
    $1.04\times$ on the TP ring's chassis.
    \item \textbf{\textsf{Baseline}:} it is not lockstepped, so each chassis pays
    only for its own load.
\end{itemize}
A $32$k-token request takes a larger share of one unit's load; on 405B it adds
only $0.04$\,ms of attention per layer. A single line or ring therefore serves the
whole context distribution.

\begin{figure}[t]
    \centering
    \includegraphics[width=1\columnwidth]{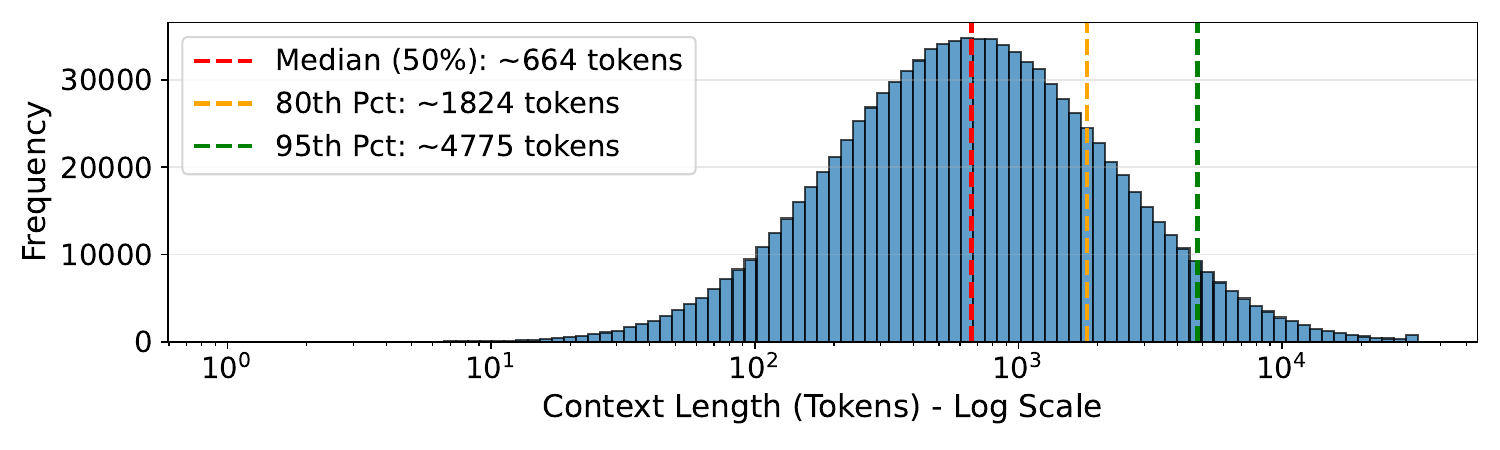}
    \caption{Distribution of request's context length at completion in the
    evaluated workload, fitted to LMSYS-Chat-1M.}
    \label{fig:context}
\end{figure}

\paragraph{Workload and cluster throughput}
Each request's total context at completion follows the log-normal of
\autoref{fig:context}, fitted to LMSYS-Chat-1M~\cite{zheng2024lmsys} 
(median $664$,
$\sigma = 1.2$, clipped at $32{,}768$, mean $1{,}356$). Its output length is
log-normal (median $200$, $\sigma = 1.0$, our assumption) and bounded by that
total, and the rest is the prompt. A decode step sees the requests in flight,
each once per output token, so KV load and attention follow the step-weighted
mean context, $1{,}450$ tokens; the placement simulation measures the same value. All
systems serve this distribution as a single pool. 
Cluster throughput is $X = 80{,}000 \cdot x$, where $x$ is the decode
throughput per accelerator (tokens/s), and TBT is the step time.

\paragraph{Power and energy}
Energy per token is cluster static power divided by $X$, plus dynamic energy per token.
\begin{itemize}
    \item \textbf{Static power.} Every accelerator draws $150$\,W at idle~\cite{onesource2026h100idle} plus
    $0.2$\,W/GB of HBM refresh ($25.6$\,W at 128\,GB). The baseline adds $44.4$\,W
    of NVLink SerDes. \Ring{} and \Line{} add $4$\,W of receiver bias, and keep
    NVLink in their TP and DP-attention layouts. The PP \Ring{} adds $13.4$\,W of handover links.
    \item \textbf{Shared optics and servers.} MC-PDFAs draw $25$\,W
    each~\cite{fiber}. The \Line{}'s weights server has a $4.1$\,kW laser
    bank~\cite{fiber}, $0.2$\,W/GB of DRAM, and spends $5$\,pJ/bit re-reading the
    model every pass.
    \item \textbf{Ring regeneration.} Each $50$\,km spool's inline amplifier
    contributes an OSNR of $39.2$\,dB, and a pod booster tapped after an amplifier
    $41.2$\,dB. Keeping $29$\,dB at the PAM4 receivers allows $9$ spools ($450$\,km)
    between $4.25$\,kW O-E-O 3R stations. A ring is as long as one loop at
    $5\,\mu$s/km: a pipeline stage for PP, a decode step for TP.
    \item \textbf{Dynamic energy.} MACs and attention at $0.35$\,pJ/op. Weight
    delivery at $40$\,pJ per HBM3e byte read, or $5.6$\,pJ per received fiber
    byte~\cite{fiber}, amortized over the tokens that use it. KV reads at
    $40$\,pJ/B, and NVLink traffic at $20$\,pJ/B.
\end{itemize}

\subsection{Results}

\begin{table*}[t]
\centering
\caption{Decode evaluation on $80{,}000$ accelerators with \emph{equal KV capacity}: baseline GPUs have 128\,GB of HBM3e, and \Ring{} and \Line{} accelerators have only the HBM the baseline has left after its weight shard (77.4, 78.0 and 22.5\,GB), so every system can hold the same KV. 
$R$ is requests per 8-accelerator chassis; every system runs at the knee of its own throughput curve (smallest $R$ within 3\% of its peak, at most $2{,}048$), subject to KV. 
\Line{} sizes each broadcast's slack for its most loaded unit ($1.02\times$ the mean); the rings fix it, at $1.035\times$ (PP) and $1.04\times$ (TP). \Ring{} (TP) assumes that cross-cable skew is tracked. 
Collectives overlap compute with $1/16$ exposed. \Ring{} (TP) splits every layer, experts included (TP+EP), across the 8 cables; \Line{} (TP) and \Line{} (DP-attn.\,+\,EP) are the split-port layouts. Under PP, DP and DP-attention, $R$ counts the requests of all 8 accelerators (e.g.\ $1{,}984 = 8\times248$). Power is total cluster power, static plus dynamic (energy per token $\times$ throughput). Ratios are relative to the HBM baseline; energy ratios below $1$ are savings. Weight DRAM is the cluster-wide memory holding model weights; HBM saved is the cluster HBM removed relative to the baseline's 10.24\,PB (minus the weights server's copy).}
\label{tab:eval_all}
\setlength{\tabcolsep}{2.6pt}
\begin{tabular}{@{}llrrrrrrrrrrr@{}}
\toprule
 & & \textbf{$R$ per} & \textbf{TBT} & \textbf{Throughput} & \textbf{Power} & \multicolumn{3}{c}{\textbf{Energy (J/token)}} & \multicolumn{2}{c}{\textbf{vs.\ Baseline}} & \textbf{Weight} & \textbf{HBM saved} \\
\cmidrule(lr){7-9}\cmidrule(lr){10-11}
\textbf{Model} & \textbf{System} & \textbf{chassis} & \textbf{(ms)} & \textbf{(M tok/s)} & \textbf{(MW)} & \textbf{Static} & \textbf{Dyn.} & \textbf{Total} & \textbf{Thr.} & \textbf{Energy} & \textbf{DRAM} & \textbf{(10.24\,PB tot.)} \\
\midrule
\multirow{5}{*}{\begin{tabular}[c]{@{}l@{}}Llama-\\3.1-405B\end{tabular}} & \textsf{Baseline} (HBM3e) TP & 248 &  22.7 & \good 109.1 & \fair 59.6 & 0.161 & 0.385 & 0.547 & -- & -- & 4.05\,PB & -- \\
 & \Ring{} (PP) & 760 & 148.7 & 51.1 & 32.2 & 0.289 & 0.342 & 0.631 & $0.47\times$ & $1.15\times$ & 0.4\,TB & 4.05\,PB (40\%) \\
 & \Ring{} (TP) & 248 & 23.7 & \fair 104.6 & \good 51.7 & 0.165 & 0.329 & 0.494 & \good $0.96\times$ & \good $0.90\times$ & 0.4\,TB & 4.05\,PB (40\%) \\
 & \Line{} (DP) & 760 & 148.2 & 51.3 & 31.2 & 0.266 & 0.342 & 0.608 & $0.47\times$ & $1.11\times$ & 0.4\,TB & 4.05\,PB (40\%) \\
 & \Line{} (TP) & 248 & 23.6 & \fair 105.2 & \good 51.9 & 0.163 & 0.329 & 0.493 & \good $0.96\times$ & \good $0.90\times$ & 0.4\,TB & 4.05\,PB (40\%) \\
\midrule
\multirow{6}{*}{\begin{tabular}[c]{@{}l@{}}Llama-\\4-Maverick\end{tabular}} & \textsf{Baseline} (HBM3e) TP+EP & 1{,}888 & 38.9 & \good 485.4 & \fair 33.4 & 0.036 & 0.033 & 0.069 & -- & -- & 4.00\,PB & -- \\
 & \Ring{} (PP) & 1{,}976 & 146.9 & 134.5 & 19.5 & 0.112 & 0.033 & 0.145 & $0.28\times$ & $2.10\times$ & 0.4\,TB & 4.00\,PB (39\%) \\
 & \Ring{} (TP+EP) & 288 & 18.9 & 152.1 & 22.0 & 0.113 & 0.032 & 0.145 & $0.31\times$ & $2.11\times$ & 0.4\,TB & 4.00\,PB (39\%) \\
 & \Line{} (DP) & 1{,}984 & 146.5 & 135.4 & 18.3 & 0.103 & 0.033 & 0.135 & $0.28\times$ & $1.97\times$ & 0.4\,TB & 4.00\,PB (39\%) \\
 & \Line{} (TP+EP) & 288 & 18.9 & 152.6 & 22.1 & 0.113 & 0.032 & 0.145 & $0.31\times$ & $2.10\times$ & 0.4\,TB & 4.00\,PB (39\%) \\
 & \Line{} (DP-attn.\,+\,EP) & 1{,}928 & 41.0 & \fair 470.7 & \good 29.1 & 0.037 & 0.025 & 0.062 & \good $0.97\times$ & \good $0.90\times$ & 0.4\,TB & 4.00\,PB (39\%) \\
\midrule
\multirow{6}{*}{\begin{tabular}[c]{@{}l@{}}DeepSeek-\\V4-Pro\end{tabular}} & \textsf{Baseline} (HBM3e) TP+EP & 1{,}216 &  39.9 & \fair 304.8 & \fair 38.2 & 0.058 & 0.068 & 0.125 & -- & -- & 8.44\,PB & -- \\
 & \Ring{} (PP) & 1{,}992 & 265.6 & 75.0 & 18.3 & 0.189 & 0.055 & 0.244 & $0.25\times$ & $1.95\times$ & 0.8\,TB & 8.44\,PB (82\%) \\
 & \Ring{} (TP+EP) & 256 & 34.1 & 75.1 & 20.8 & 0.219 & 0.058 & 0.277 & $0.25\times$ & $2.21\times$ & 0.8\,TB & 8.44\,PB (82\%) \\
 & \Line{} (DP) & 1{,}992 & 265.4 & 75.1 & 17.1 & 0.173 & 0.055 & 0.228 & $0.25\times$ & $1.82\times$ & 0.8\,TB & 8.44\,PB (82\%) \\
 & \Line{} (TP+EP) & 256 & 34.1 & 75.1 & 20.7 & 0.217 & 0.058 & 0.276 & $0.25\times$ & $2.20\times$ & 0.8\,TB & 8.44\,PB (82\%) \\
 & \Line{} (DP-attn.\,+\,EP) & 1{,}984 & 41.4 & \good 478.8 & \good 35.0 & 0.034 & 0.039 & 0.073 & \good $1.57\times$ & \good $0.58\times$ & 0.8\,TB & 8.44\,PB (82\%) \\
\bottomrule
\end{tabular}
\end{table*}

\autoref{tab:eval_all} reports results under equal KV capacity, while \autoref{tab:eval_128} gives the comparison when every accelerator retains a full $128$\,GB of HBM. Performance is governed primarily by how each execution model partitions weights across the $N=8$ accelerators of a chassis (\autoref{tab:execmodels}).

\paragraph{Whole-model streams (\Ring{} PP and \Line{} DP)}
Under PP and DP, each accelerator ingests the entire model through a single $3.2$\,TB/s port, bounding step time from below by the full-model stream duration ($126.6$\,ms for 405B, $125.0$\,ms for Maverick, and $264.0$\,ms for V4-Pro) plus attention slack. Reaching the compute knee requires $R \approx N b^*$ requests per chassis (\autoref{tab:execmodels}). On 405B with equal KV ($77.4$\,GB per accelerator), local memory caps $b$ at $95$ ($R=760$), limiting throughput to $0.47\times$ the baseline at $1.11\times$ (\Line{}) and $1.15\times$ (\Ring{}) the energy per token. On Maverick and V4-Pro, streaming all routed experts to every accelerator restricts throughput to $0.28\times$ and $0.25\times$ the baseline. Compared to \Line{} (DP), \Ring{} (PP) incurs slightly higher latency and $4$--$7\%$ more energy per token due to pipeline handovers, higher static slack ($1.035\times$ vs.\ $1.02\times$ the mean load), and multi-thousand-kilometer loop regeneration overhead.

\paragraph{Tensor-parallel \Ring{} and \Line{}}
Splitting every layer across the $8$ MCF cables shortens the weight stream per accelerator by $8\times$ ($15.8$\,ms for 405B). Assuming cross-cable skew is actively corrected across all $2L$ all-reduces per token, \Ring{} (TP) and \Line{} (TP) both achieve $0.96\times$ the baseline's throughput on 405B at $0.90\times$ its energy per token ($23.7$ and $23.6$\,ms TBT vs.\ $22.7$\,ms on the baseline). However, TP+EP fails to scale on MoE models for both topologies: because all $8$ accelerators apply their slice of shared and attention weights to the same $R$ tokens, the per-byte knee ($b^* = 263$) caps the entire chassis at $R = b = 288$ (Maverick) and $256$ (V4-Pro) requests,%
 compared to $1{,}888$ and $1{,}216$ on the baseline. Consequently, TP+EP achieves lower TBT ($18.9$ and $34.1$\,ms) but only $0.31\times$ and $0.25\times$ the baseline's throughput at $2.10$--$2.21\times$ the energy. A ring could escape this bottleneck only by carrying the shared weights on all $8$ cables, as the split-port \Line{} does below; for Maverick, that lengthens the loop to  $\sim 8{,}300$\,km ($2.2\times$ PP), with proportionally more amplifiers and 3R stations, for no gain in throughput over \Line{}.

\paragraph{Split-port \Line{} (DP-attn.\,+\,EP)}
Because an open-ended line has no loop-capacity constraint, \Line{} can pair data-parallel attention with expert parallelism (DP-attn.\,+\,EP) for MoE models, broadcasting shared weights on all $8$ cables while partitioning experts $8$ ways. Each accelerator applies shared weights only to its own $b = R/8 \approx 241$--$248$ requests, allowing the chassis to serve $R = 1{,}928$ (Maverick) and $1{,}984$ (V4-Pro) requests without exceeding $b^*$.

With equal KV, split-port \Line{} (TP on 405B, DP-attn.\,+\,EP on Maverick) achieves $0.96\times$ and $0.97\times$ the baseline's throughput at $0.90\times$ the energy per token, while eliminating $4.05$\,PB ($40\%$) and $4.00$\,PB ($39\%$) of cluster HBM. Their TBT stays within $4$--$5\%$ of the baseline ($23.6$ vs.\ $22.7$\,ms on 405B; $41.0$ vs.\ $38.9$\,ms on Maverick), reflecting the $4\%$ bandwidth difference ($3.2$ vs.\ $3.35$\,TB/s) and $1.02\times$ lockstep slack. Energy savings stem from replacing $40$\,pJ/B HBM weight reads with $5.6$\,pJ/B CPO reception and eliminating weight-HBM refresh power.

On DeepSeek-V4-Pro, weights ($105.5$\,GB/GPU) crowd out KV capacity in the $128$\,GB baseline, and its single KV head forces the TP+EP baseline to replicate KV state across all $8$ GPUs in a chassis, capping load at $R = 1{,}216$. Because \Line{} (DP-attn.\,+\,EP) stores no weights locally and partitions requests across accelerators without KV replication, it uses only $4.2$\,GB of its $22.5$\,GB HBM budget to reach the compute knee ($R = 1{,}984$), delivering $1.57\times$ the throughput at $0.58\times$ the energy per token ($41.4$ vs.\ $39.9$\,ms TBT) while saving $8.44$\,PB ($82\%$) of cluster HBM.
   The baseline cannot adopt this layout within a chassis: each GPU would have
   to store the shared weights and an eighth of the experts
   ($26 + 819/8 \approx 128$\,GB), leaving no room for KV.

\begin{table}[t]
\centering
\caption{Results with 128\,GB of HBM: throughput and energy per token relative
to the baseline's absolute values (M tokens/s, J/token).
TP rows are TP+EP for the MoE models; DP-attention with EP runs only on
\Line{}: on V4-Pro, the baseline cannot fit this layout's weights
($\approx 128$\,GB) and KV.}
\label{tab:eval_128}
\scriptsize
\setlength{\tabcolsep}{2.5pt}
\begin{tabular}{@{}lrrrrrr@{}}
\toprule
 & \multicolumn{2}{c}{\textbf{405B}} & \multicolumn{2}{c}{\textbf{Maverick}} & \multicolumn{2}{c}{\textbf{V4-Pro}} \\
\cmidrule(lr){2-3}\cmidrule(lr){4-5}\cmidrule(lr){6-7}
\textbf{System} & \textbf{Thr.} & \textbf{Energy} & \textbf{Thr.} & \textbf{Energy} & \textbf{Thr.} & \textbf{Energy} \\
\midrule
\textsf{Baseline} (abs.) & 109.1 & 0.547 & 485.4 & 0.069 & 304.8 & 0.125 \\
\midrule
\Ring{} (PP) & $0.70\times$ & $0.98\times$ & $0.28\times$ & $2.19\times$ & $0.25\times$ & $2.12\times$ \\
\Ring{} (TP) & \good $0.96\times$ & \good $0.92\times$ & $0.31\times$ & $2.19\times$ & $0.25\times$ & $2.39\times$ \\
\Line{} (DP) & $0.71\times$ & $0.95\times$ & $0.28\times$ & $2.05\times$ & $0.25\times$ & $2.00\times$ \\
\Line{} (TP) & \good $0.96\times$ & \good $0.92\times$ & $0.31\times$ & $2.18\times$ & $0.25\times$ & $2.38\times$ \\
\Line{} (DP-attn.\,+\,EP) & -- & -- & \good $0.97\times$ & \good $0.92\times$ & \good $1.57\times$ & \good $0.61\times$ \\
\bottomrule
\end{tabular}
\end{table}

\paragraph{128\,GB per accelerator and iso-throughput scaling}
When every accelerator keeps $128$\,GB of HBM (\autoref{tab:eval_128}), only the KV-constrained 405B whole-model streams improve, rising from $0.47\times$ to $0.70\times$ (\Ring{} PP) and $0.71\times$ (\Line{} DP) relative throughput. Split-port layouts already operate at the knee, so their throughput is unchanged while unused HBM refresh slightly shifts their relative energy to $0.92\times$ (405B, Maverick) and $0.61\times$ (V4-Pro).

At equal throughput with equal KV, 
split-port \Line{} reduces cluster power by $10\%$ on 405B ($53.8$ vs.\ $59.6$\,MW) and Maverick ($30.0$ vs.\ $33.4$\,MW) using $3$--$4\%$ more accelerators, and by $42\%$ on V4-Pro ($22.3$ vs.\ $38.2$\,MW) using $36\%$ fewer accelerators ($50.9$k vs.\ $80.0$k). \Line{} delivers $1.59\times$, $1.59\times$ and $8.9\times$ the baseline's tokens/s per TB of HBM ($8\%$, $8\%$ and $39\%$ power savings when retaining $128$\,GB per accelerator).

\begin{HIDDEN}
\begin{table}[t]
\centering
\caption{Cluster power (MW) and accelerators needed to deliver the baseline's decode throughput, and decode tokens/s per TB of HBM, for the HBM baseline and the split-port \Line{}: TP for 405B, and DP-attention with experts split across ports (EP) for the MoE models. \emph{128\,GB}: every accelerator has 128\,GB, as in \autoref{tab:eval_128}. \emph{Equal KV}: \Line{} accelerators carry only the HBM the baseline has left after its weight shard (77.4, 78.0 and 22.5\,GB), so both hold the same KV, as in \autoref{tab:eval_all}.}
\label{tab:eval_iso}
\scriptsize
\setlength{\tabcolsep}{1.2pt}
\begin{tabular}{@{}llrrrrrr@{}}
\toprule
 & & \multicolumn{3}{c}{\textbf{128\,GB}} & \multicolumn{3}{c}{\textbf{Equal KV}} \\
\cmidrule(lr){3-5}\cmidrule(lr){6-8}
\textbf{Model} & & \textbf{MW} & \textbf{Accel.} & \textbf{Tok/s/TB} & \textbf{MW} & \textbf{Accel.} & \textbf{Tok/s/TB} \\
\midrule
\multirow{2}{*}{405B} & \textsf{Base} & 59.6 & 80.0k & 10.7k & 59.6 & 80.0k & 10.7k \\
 & \Line{} (TP) & 54.6 & 83.0k & 10.3k & 53.8 & 83.0k & 17.0k \\
\midrule
\multirow{2}{*}{Maverick} & \textsf{Base} & 33.4 & 80.0k & 47.4k & 33.4 & 80.0k & 47.4k \\
 & \Line{} (EP) & 30.9 & 82.5k & 46.0k & 30.0 & 82.5k & 75.4k \\
\midrule
\multirow{2}{*}{V4-Pro} & \textsf{Base} & 38.2 & 80.0k & 29.8k & 38.2 & 80.0k & 29.8k \\
 & \Line{} (EP) & 23.4 & 50.9k & 46.8k & 22.3 & 50.9k & 266.3k \\
\bottomrule
\end{tabular}
\end{table}
\end{HIDDEN}

\paragraph{Expert skew}
Whole-model \Line{} (DP) is immune to expert routing skew because every accelerator receives all experts locally. In the baseline, routing $80\%$ of Maverick's $1{,}888$ tokens to a single hot expert without replication would add $5.4$\,ms of GEMM and $1.9$\,ms of all-to-all time to a $38.9$\,ms step. Under split-port EP, the same skew would send about $1{,}540$ of the chassis's $1{,}928$ tokens to one expert, far above $b^* = 263$, so the accelerator holding it would miss weights. The solution is for the \Line{} weights server to dynamically replicate hot experts across all ports. We have not studied this as it is an on-line optimization and we leave it for future work.

\paragraph{Sensitivity}
Across sensitivity sweeps with equal KV, load-balancing quality has the largest impact because slack must cover the most loaded unit: relaxing placement to power-of-two choices ($1.26\times$ reserve) or random placement ($1.9\times$) changes split-port \Line{} relative throughput to $0.90\times$/$0.77\times$ (405B), $0.85\times$/$0.68\times$ (Maverick), and $1.79\times$/$2.50\times$ (V4-Pro, where higher reserves further starve the baseline's scarce KV space). Lowering GEMM efficiency to $\eta = 0.6$ ($b^* = 186$) shifts relative throughput to $0.99\times$, $0.92\times$, and $1.19\times$, while provisioning $192$\,GB per baseline GPU yields $0.96\times$, $0.97\times$, and $0.96\times$ throughput at $0.90\times$, $0.88\times$, and $0.80\times$ energy. Uncorrected $10$\,K cross-cable thermal drift ($1.8\,\mu$s per collective) degrades \Ring{} (TP) on 405B from $0.96\times$ to $0.94\times$, and varying collective exposure ($0$ to $1/4$), knee thresholds ($90\%$), or median output lengths ($400$ tokens) alters relative throughput by at most $0.04\times$. If the log-normal is instead read as the prompt length, with outputs appended (step-weighted mean $1{,}784$), split-port \Line{} reaches $0.97\times$, $0.95\times$ and $1.71\times$ the baseline's throughput, and the KV-bound whole-model streams fall to $0.41\times$ on 405B.

\begin{HIDDEN}
\section{Related Work}

\paragraph{Optical Interconnects for ML} 
Silicon photonics (SiP) has gained significant traction for scaling ML clusters. Systems like SiP-ML~\cite{sip-ml} utilize co-packaged optics (CPO) to replace electrical scale-up networks, and recent proposals like Fiber Memory~\cite{fiber} suggested using optical loops as physical memory delay-lines. Our work re-imagines the asynchronous delay-line paradigm, replacing the fiber ring with a tree topology. 

\paragraph{LLM Inference Systems}
State-of-the-art inference engines like vLLM~\cite{vllm} rely heavily on PagedAttention to maximize KV cache utilization within constrained local HBM. By removing weights from the local HBM, our architecture provides a massive expansion of available KV cache capacity per node, natively complementing continuous batching systems.

\end{HIDDEN}

\section{Conclusion}

Optical Broadcast Memory (\Line{}) shows that the read-only weights of
datacenter-scale LLM inference need not be replicated in the HBM of every
accelerator. A single weights server broadcasts the model over an open-ended,
hierarchical Tap-and-Amplify line that receive-only co-packaged optics feed
directly into activation-stationary accelerators.
At datacenter scale, this frees \textbf{petabytes} of expensive,
supply-constrained HBM for the KV cache, or removes it altogether.

Our evaluation yields four insights. First, the execution model matters more
than the optics: streaming the whole model to every accelerator is bound by
per-accelerator fiber bandwidth, while splitting the ports of a chassis across
layers or experts matches an HBM baseline. Second, because a push stream cannot
be paused, the compute knee is a hard limit for every weight byte, so weights
shared by all tokens and not the routed experts, bound the load of MoE models.
Third, \Line{}'s freedom from the \Ring{}'s fixed loop length is what makes the
DP-attention layout for MoE practical, and lets it size slack for the current
load at every broadcast rather than fix it, as the Ring must, for the worst
load it will see.
Fourth, the benefits come from cheaper weight
delivery and freed memory, not from extra bandwidth, and they are largest where
weights crowd out the KV cache. Since fiber bandwidth is not tied to the memory
stacks on the package, provisioning each accelerator with the bandwidth its
compute requires is a natural next step toward decoupling model size from
accelerator memory.

\bibliographystyle{IEEEtranS}
\bibliography{references}

\end{document}